\documentclass{optica-article}

\journal{opticajournal} % for journals or Optica Open

\articletype{Research Article}

\usepackage{dcolumn}% Align table columns on decimal point
\usepackage{bm}
\usepackage{graphicx} % Required for inserting images
\usepackage{chemformula} % for... Chemical formula
\usepackage[uncertainty-mode = separate, separate-uncertainty-units = single]{siunitx}
\usepackage{xcolor}
\usepackage{float}
\usepackage{amsmath,amsfonts}
\usepackage{acronym}
\usepackage{stfloats}

\definecolor{wong_blue}{RGB}{0, 114, 178}
\definecolor{wong_orange}{RGB}{230, 159, 0}
\definecolor{wong_skyblue}{RGB}{86, 180, 233}
\definecolor{wong_bluishgreen}{RGB}{0, 158, 115}
\definecolor{wong_yellow}{RGB}{240, 228, 66}
\definecolor{wong_vermillon}{RGB}{213, 94, 0}
\definecolor{wong_reddishpurple}{RGB}{204, 121, 167}
\definecolor{grey}{gray}{0.2}
\definecolor{color_gqd_solo}{RGB}{0, 114, 178}
\definecolor{color_mapb}{RGB}{0, 101, 0}
\definecolor{color_trpl_high_energy}{RGB}{0, 158, 115}
\definecolor{color_trpl_low_energy}{RGB}{230, 159, 0}
\definecolor{color_pl_gqd_solution}{RGB}{213, 94, 0}
\definecolor{color_trpl_gqd_solution}{RGB}{213, 94, 0}
\definecolor{color_ple_gqd_solution}{RGB}{148, 0, 166}
\definecolor{color_abs_gqd_solution}{RGB}{148, 0, 166}
\definecolor{color_ple_gqd_poly}{RGB}{0, 114, 178}
\definecolor{color_trpl_gqd_poly}{RGB}{0, 114, 178}
\definecolor{color_ple_gqd_mapb}{RGB}{0, 38, 61}
\definecolor{color_trpl_samples_1}{RGB}{204, 121, 167}
\definecolor{color_trpl_samples_2}{RGB}{86, 180, 233}
\definecolor{color_trpl_all_energies}{RGB}{0, 0, 0}

\usepackage{hyperref}

\usepackage{xr}
\usepackage{subcaption}
\DeclareCaptionLabelFormat{bf-parens}{(\textbf{#2})}
\newcommand{\labfig}[1]{\label{fig:#1}}
\newcommand{\reffig}[1]{\hyperref[fig:#1]{\figurename}~\ref{fig:#1}}
\newcommand{\refsupfig}[1]{Supp. \figurename~\ref{fig:#1}}

\newcommand{\gqd}{\ch{C96\textit{t}Bu8}}
\newcommand{\mapb}{\ch{MAPbBr3}}
\newcommand{\subreffig}[1]{\protect\subref{fig:#1}}

\newacro{GQD}{Graphene Quantum Dot}
\newacro{h-BN}{hexagonal boron nitride}
\newacro{ZPL}{Zero-Phonon Line}
\newacro{TRPL}{Time-Resolved Photo Luminescence}
\newacro{IRF}{Instrument Response Function}
\newacro{TMDCs}{transition-metal dichalcogenides}
\newacro{TTTR}{Time-Tagged Time-Resolved}
\newacro{THF}{tetrahydrofuran}
\newacro{PL}{photoluminescence}
\newacro{PDI}{perylenediimide}

\begin{document}

\title{Collective Fluorescence of Graphene Quantum Dots on a surface}

\author{Hugo Levy-Falk,\authormark{1,2,$\dag$} Suman Sarkar,\authormark{1,$\dag$} Thanh Trung Huynh,\authormark{1} Daniel Medina-Lopez,\authormark{3} Lauren Hurley,\authormark{1} Océane Capelle,\authormark{1} Muriel Bouttemy,\authormark{4} Gaëlle Trippé-Allard,\authormark{1} Stéphane Campidelli,\authormark{3} Loïc Rondin,\authormark{1} Elsa Cassette,\authormark{1} Emmanuelle Deleporte,\authormark{1} Jean-Sébastien Lauret\authormark{1,*}}

\address{\authormark{1}Laboratoire LuMIn, Université Paris-Saclay, ENS Paris-Saclay, CentraleSupélec, CNRS, LuMIn, 91190 Gif-sur-Yvette, France\\
\authormark{2}National Institute of Optics (CNR-INO), c/o LENS via Nello Carrara 1, Sesto F.no 50019, Italy\\
\authormark{3}Université Paris Saclay, CEA, NIMBE, LICSEN, \textsc{Gif-sur-Yvette, France}\\
\authormark{4}Institut Lavoisier de Versailles, UVSQ, Université Paris-Saclay, CNRS, UMR 8180, 45 avenue des Etats-Unis, 78035 Versailles Cedex, France\\
\authormark{$\dag$}The authors contributed equally to this work.}

\email{\authormark{*}lauret@ens-paris-saclay.fr}

\begin{abstract*}
    This study explores the organization of graphene quantum dots on the surface of monocrystalline halide perovskite. We show that graphene quantum dots tends to aggregate on the surface of perovskite unlike in solution or on other substrates, even at very low concentration of the initial solution that should yield single-molecule samples.  Spectral analysis on small clusters shows a back-and-forth dynamical transition between an uncoupled,  monomer-like state, and an excimer state. Following this "dance" between states, a drastic one-way increase in fluorescence intensity combined with a shortening of the excited state lifetime has been observed on some clusters.  This behavior is related to the emission of a collective state that may be a consequence of the dynamical organization of graphene quantum dots under illumination on the surface of the perovksite.
\end{abstract*}

\section{Introduction}

Light-matter interaction is at the heart of many applications, such as optoelectronics, photovoltaics, and quantum technologies. In this perspective, new emitting materials are constantly being developed. One way to tailor the optical properties is to combine two or more materials to build heterostructures. It can lead to physical processes such as charge and energy transfer or proximity effects \cite{proximity, transfer1, transfer2, cyrielle}. These effects have been studied extensively in heterostructures of 2D materials such as graphene, \ac{h-BN}, or \ac{TMDCs}. The coupling between molecular and solid-state materials shows some assets in this context. Organic chemistry offers a wide choice of molecules whose properties can be tuned at will. For instance, phthalocyanine molecules were coupled to TMDCs to modulate their optoelectronic properties through photoinduced charge transfer~\cite{choi16}. Likewise, including perylene-3,4,9,10-tetracarboxylic diimide (PTCDI) molecules in a graphene/\ac{h-BN} heterostructure made it possible to control their triplet state electrically~\cite{sva20}. As another example, N,N'-dimethyl-3,4,9,10-perylentetracarboxylicdiimide (MePTCDI) molecules on the surface of \ac{h-BN} tend to self-organize in 2D, leading to interesting collective states showing redshifted and narrower optical transitions in comparison to the monomer~\cite{juergensenCollectiveStatesMolecular2023, collectiveMePTCDI}. In all these studies, the organization of the molecules on the surface of the solid plays a crucial role. 

Here, we study the organization of recently reported elongated \ac{GQD}~\cite{medina-lopezInterplayStructurePhotophysics2023} on the surface of a hybrid organic-inorganic perovskite single crystal (\ch{CH3NH3PbBr3}, called \mapb{}). These \ac{GQD}s, synthesized by bottom-up chemistry, are highly soluble in \ac{THF} and show a high fluorescence quantum yield ($>90\%$ )~\cite{medina-lopezInterplayStructurePhotophysics2023}. Moreover, their optical bandgap can be tuned over more than 100~nm by increasing in a controlled way the number of sp$^2$ carbon atoms from 78 to 132~\cite{medina-lopezInterplayStructurePhotophysics2023}. These properties make them promising materials for light-emitting devices \cite{OLED}, in particular as single quantum emitters~\cite{levy-falkInvestigationRodShapedSingleGraphene2023a, Zhao2018}. On the other hand, \mapb{} halide perovskite is a direct bandgap semiconductor that absorbs and emits light in the visible range, and is easy to grow by solution chemistry. It is widely studied in the context of photovoltaic~\cite{Zhu2022} and light-emitting applications such as LEDs~\cite{Kim2025} and lasers~\cite{bouteyreDirectingRandomLasing2020, Afify2022}. The combination of \ac{GQD}s and \mapb{} is enticing to develop new functionalities through proximity effects. In particular, \ac{GQD}s, being made only of carbon atoms, show low spin-orbit coupling. On the contrary, lead-based perovskites show strong spin-orbit coupling~\cite{spin-orbit} that could be transferred to \ac{GQD}s through proximity effects. This could allow tuning the singlet to triplet intersystem crossing in \ac{GQD}, enabling their use in spin devices. Efficient transfer of spin-orbit coupling has already been reported for monolayer graphene in contact with few-layer tungsten disulfide \cite{proximity2}. Before addressing these effects, the investigation of the organization of \ac{GQD}s on the surface of MAPB is mandatory. Here, we show that, in contrast to their behavior in solution and on other substrates, the \ac{GQD}s form clusters when deposited on the surface of \mapb{} crystal, whatever the concentration of the initial solution of \ac{GQD}s. In particular, when the concentration is decreased so as to observe diffraction-limited spots in the microphotoluminescence of \ac{GQD}s, we show that the emission of these clusters undergoes dynamical spectral jumps characteristic of a transition between coupled and uncoupled molecular states. Moreover, these jumps are sometimes further followed by a stabilization to a final state, showing a drastic increase in the \ac{PL} intensity and a reduced excited state lifetime, characteristic of a collective state.

\section{Experimental Section}
\label{sec:methods}

\begin{figure}
    \centering
     \begin{subcaptiongroup}
        \subcaptionlistentry{}
        \labfig{molecule gqd}
        \subcaptionlistentry{}
        \labfig{pl poly gqd}
    \end{subcaptiongroup}
    \includegraphics{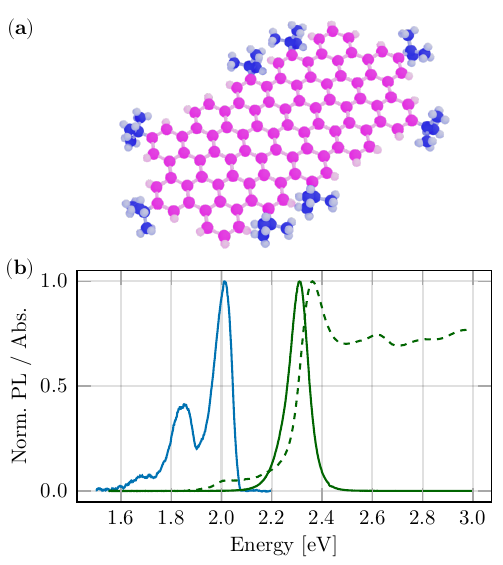}
    \caption{\subreffig{molecule gqd}~Schematic representation of \gqd{}, with the bulky \textit{tert}-butyl groups (\textcolor{wong_blue}{blue}). \subreffig{pl poly gqd}~Typical optical properties of the \ac{GQD}s and \mapb{} perovskite: Emission spectrum of a single \gqd{} measured in a polystyrene matrix (\textcolor{color_gqd_solo}{blue}). Absorption (\textcolor{color_mapb}{dashed green line}) and emission (\textcolor{color_mapb}{solid green line}) spectra of the \mapb{} substrate. The single molecule emission spectrum and the perovskite absorption spectrum have been smoothed to represent their typical shape.}
    \labfig{presentation of gqd}
\end{figure}

\subsection*{Synthesis and characterization of \mapb{} crystals}
A solution at \SI{0.9}{M} of \ch{MAPbBr3} with a 1.1 excess of \ch{MABr} was prepared by dissolving \SI{111}{mg} of \ch{MABr} and \SI{331}{mg} of \ch{PbBr2} in \SI{1}{mL} of N, N-dimethylformamide (DMF). The synthesis of \ch{MABr} is described in the supplementary materials of Ref.~\cite{bouteyreRoomTemperatureCavityPolaritons2019}. The solution is agitated in an ultrasound bath at \SI{50}{\degreeCelsius} until it is clear and transparent. A droplet (typically $\sim\SI{10}{\micro\L}$) is then deposited between two, pre-cleaned with O$_2$ plasma, thin glass coverslips and left to crystallize at room temperature for a few days. The glass coverslips are thin enough to be easily separated once the crystallization process is over by peeling one of them. The crystals are expected to inherit the flatness of the covering glass. They are typically tens of micrometers thick and up to a millimeter large, as shown in \refsupfig{optical microscope}. The surface topography and composition of the crystals has been investigated using atomic force microscopy (AFM) and X-ray photo-emission spectroscopy (XPS). In particular, the former showed a $\sim\SI{10}{nm}$ roughness (see \refsupfig{afm}). Additionally, the latter demonstrated that \mapb{} crystals present a relatively clean surface with some metallic lead and low carbon contamination at the surface (see \refsupfig{xps big} and \refsupfig{xps}), making it suitable for our purposes. Control experiments performed on the bare substrate, where the sample was excited at 594 nm ($\sim$~2.09~eV), wavelength used to excite the \ac{GQD}, show no \ac{PL} signal. This is consistent with the absorption threshold of \mapb{} ($\sim$~2.2eV) as shown on \reffig{pl poly gqd}. Here, the absorption spectrum has been measured on a spin-coated thin film of \mapb{} because the single crystals cannot be measured directly as they are too thick. However, the emission spectrum of the single crystal can be measured with an excitation above the bandgap at 3.06~eV, as shown in \reffig{presentation of gqd}.

\subsection*{Synthesis of GQDs}

The synthesis of the rod-shaped \gqd{} \ac{GQD} is based on the Scholl oxidation of the corresponding polyphenylene dendrimer (denoted 1 in \refsupfig{synthesis oxydation}) with 2,3-dichloro-5,6-dicyano-1,4-benzoquinone (DDQ) in the presence of triflic acid in \ch{CH2Cl2}. The polyphenylene dendrimer (1) was synthesized following the procedure available in our previous report~\cite{medina-lopezInterplayStructurePhotophysics2023}.

\subsection*{Deposition of GQDs on the substrate}

Solutions containing \gqd{} were deposited on freshly peeled \mapb{} substrates by spin-coating. The typical concentration of the solution was $\sim$\SI{0.15}{\micro M} in \ac{THF}. The solution was spin-coated at \SI{1000}{rpm}, with an acceleration of \SI{100}{rpm/s} and rotation duration of \SI{180}{s}.

\subsection*{Confocal fluorescence microscopy}
The experiments were conducted at room temperature on a homemade confocal fluorescence microscope. A schematic representation of the setup is given in \refsupfig{confocal microscope}. The samples were excited either in close resonance to the \ac{ZPL} of \gqd{}, using a \SI{594}{nm} ($\sim$\SI{2.09}{eV}) continuous-wave laser (Cobolt Mambo 100) or a \SI{60}{MHz} pulsed Fianium supercontinuum laser (excitation at $\sim$\SI{570}{nm}, $\sim$\SI{2.18}{eV}). The laser beam was focused on the sample through a microscope objective (\SI{0.95}{NA}, MPLAPON100X M Plan Apochromat from Evident Scientific). The \ac{PL} was collected through the same objective and separated from the excitation beam by a laser beam splitter (zt 594 RDC, Chroma). Residual light from the laser was filtered using a long-pass filter (FELH0600, Thorlabs). Photon detection was performed using silicon avalanche photodiodes (SPCM-AQR-13, Excelitas), and emission spectra were recorded using a spectrograph (SP-2350, Princeton Instruments) and a liquid nitrogen-cooled charge-coupled device camera (PyLoN:100Br eXcelon, Princeton Instruments). Raster scans were performed using a PCIe-6323 (National Instruments) card for controlling the piezo stage of the microscope and counting the photon events in the Qudi framework~\cite{binderQudiModularPython2017}. Photon statistics and time-tagging experiments were performed using a PicoHarp 300 (PicoQuant) card. The \ac{TTTR} measurements are analyzed in sections of one second and treated as \ac{TRPL} measurements. Because the repetition rate of the Fianium laser is not low enough compared to the de-excitation rate of \gqd{}, we use an \ac{IRF} that consists of a laser pulse and a prepulse, as illustrated in \refsupfig{explanation fitting procedure}. It allows accounting for the overlap of photons emitted from the pulse preceding the laser pulse -- which also triggers the detection -- in the fitting procedure.

\subsection*{Data Analysis and Plotting}
The data analyses and simulations were performed using the Julia programming language~\cite{bezansonJuliaFreshApproach2017} and specialized libraries~\cite{DrWatson2020, JSSv098i16, Measurements.jl-2016, JSSv107i04}. The Makie library~\cite{DanischKrumbiegel2021} was used for the plots.

\section{Results and Discussion}

\begin{figure}
    \centering
    \begin{subcaptiongroup}
        \subcaptionlistentry{PL vs concentration map}
        \labfig{PL vs concentration map}
        \subcaptionlistentry{PL vs concentration spectra}
        \labfig{PL vs concentration spectra}
    \end{subcaptiongroup}
    \includegraphics[width=\linewidth]{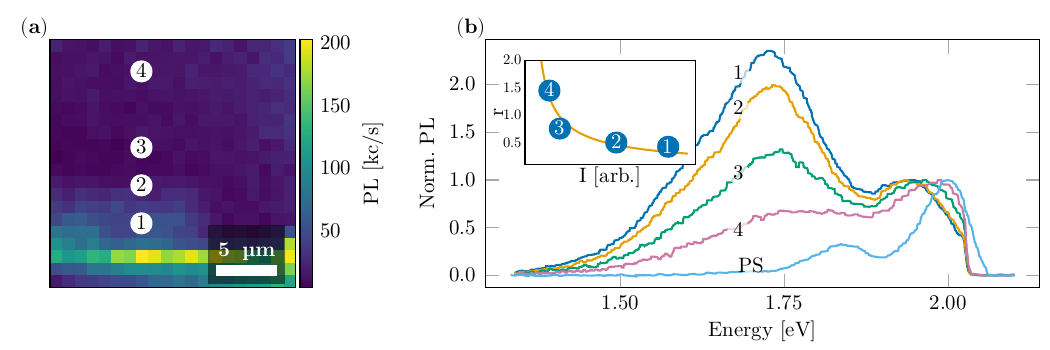}
    \caption{\subreffig{PL vs concentration map}~PL raster scan of \gqd{} films on \mapb{} crystal, the brighter points indicate higher integrated PL and thus higher concentration of emitters. \subreffig{PL vs concentration spectra}~PL spectra taken on points indicated in \subreffig{PL vs concentration map}. A typical spectrum in a polystyrene matrix is plotted in light cyan for reference. The inset shows the ratio $r$ between the peak at high energy and the peak at lower energy. The yellow line is a guide to the eye and corresponds to a power law fit of the ratio against intensity, with characteristic exponent $-0.62$.}
    \labfig{pl vs concentration}
\end{figure}

The chemical structure of \gqd{} \ac{GQD} is shown on \reffig{molecule gqd}. A typical \ac{PL} spectrum of a single \gqd{} in a polystyrene matrix at room temperature is displayed on \reffig{pl poly gqd} in blue. Second-order correlation measurements have been performed to ensure that a single \gqd{} is under investigation (see \refsupfig{g2 ps}). The emission spectrum is typical of polycyclic aromatic hydrocarbon at room temperature, with a Zero-Phonon Line (\ac{ZPL}) centered at approximately \qty{2.00(0.01)}{\eV} (FWHM \qty{92(8)}{\meV}), and its vibronic replica at \qty{1.838(0.008)}{\eV} (0-1) and \qty{1.67(0.03)}{\eV} (0-2). The error estimation for the polystyrene matrix is given by statistics measured on 375 single molecules. The measured spacing of \qty{171(15)}{\meV} and \qty{167(3)}{\meV} between the \ac{ZPL} and the first vibronic replica, and the first and the second vibronic replica, respectively, is consistent with the energy of the C=C streching mode~\cite{Zhao2018}. On the other hand, the emission peak of \mapb{} lies at \SI{2.31}{eV}. The absorption spectrum is typical of \mapb{}~\cite{Zuo2017, chenElucidatingPhaseTransitions2018, bouteyreRoomTemperatureCavityPolaritons2019}, with an excitonic peak at \SI{2.35}{eV}. The absorption energy of the exciton peak of \mapb{} is sufficiently higher than that of the \ac{GQD}, such that it is possible to excite the \ac{GQD}s at 2.09~eV without exciting the perovskite.

First, we drop-casted solutions containing \gqd{}s on \mapb{} crystals. We observe the fluorescence of films of molecules (\refsupfig{film on mapb}), where the \ac{GQD}s aggregate over the geometric features of the crystal (\refsupfig{map mapb}). We also performed control experiments where only the organic solvent was dropcasted on the surface of \mapb{}. It demonstrates that the observed fluorescence does arise from \ac{GQD}s (\refsupfig{control mapb}). By contrast, \gqd{}s dropcasted on glass tends to form homogeneous films (\refsupfig{film on glass}). If the film of \ac{GQD}s is more homogenous on glass, it turns out that the fluorescence is much more stable on \mapb{} (\refsupfig{stability}). This stability of the emission makes \mapb{} a good substrate for \ac{GQD}s. The confocal photoluminescence map reported on \reffig{pl vs concentration}(a) shows an inhomogeneous distribution of the fluorescence of \ac{GQD}s on the surface of \mapb{}. The variation of fluorescence intensity is related to the inhomogeneous concentration of \ac{GQD}s on the surface. \reffig{pl vs concentration}(b) displays the evolution of the emission spectra for different concentrations of \ac{GQD}s.  For comparison purposes, the emission of a single \ac{GQD} in polystyrene is added to the figure. First, the \ac{ZPL} line of \ac{GQD}s on \mapb{} is slightly redshifted (\SI{20}{meV}) in comparison with the polystyrene case, probably because of the different dielectric environment. Moreover, the comparison between the spectrum with the lowest concentration of \ac{GQD}s and the one in polystyrene shows a reduction of the intensity ratio between the \ac{ZPL} and the 0-1 emission lines. Likewise, as the concentration of \ac{GQD}s increases, the 0-1 line becomes more and more intense as compared to that of \ac{ZPL}. The evolution of this ratio with the integrated intensity is plotted in the inset of \reffig{pl vs concentration}(b). It shows a continuous decrease with the concentration of \ac{GQD}s on the surface. Therefore, the modification of the spectrum shape can be attributed to the interaction between an increasing number of \ac{GQD}s as they aggregate on the surface. In contrast with the modification of the fluorescence spectrum, the excitation spectrum of the \ac{GQD} film on \mapb{} is pretty comparable to the ones in solution and in polystyrene. These observations are compatible with an excimer-like coupling between \ac{GQD}s aggregated on the surface. Indeed, an excimer-like emisison is related to the back to equilibirum relaxation of an excited dimer arising from the coupling between a monomer in its ground state and one in the excited state.
Moreover, in situations where the distance between molecules is of the order of the electronic wavefunction extension, the point dipole approximation fails to describe the coupling between molecules. In these situations, Spano and co-workers included intra- and inter-molecular vibronic coupling and charge-transfer processes in order to describe the modification of the molecule fluorescence spectrum~\cite{hestandExpandedTheoryJMolecular2018, bialasHolsteinPeierlsApproach2022}. In their calculations, they show that the emission in the \ac{ZPL} can be suppressed, leading to an emission spectrum in the form of a vibronic progression where the main line corresponds to the 0-1 transition~\cite {hestandExpandedTheoryJMolecular2018}. It corresponds well to our observations.

In a second step, we tried to investigate the properties of single \ac{GQD}s on the \mapb{} substrate. To do so, we decreased the concentration of \ac{GQD}s, used a more volatile solvent (\ac{THF}), and used spin-coating rather than drop-casting (\refsupfig{substrate and film characterization}) for the deposition. It results in the observation of diffraction-limited spots in \ac{PL} raster scans (see \reffig{confocal raster scan}). The spots are uniformly distributed on the surface of the perovskite substrate, similar to what is usually observed in the polystyrene matrix~\cite{levy-falkInvestigationRodShapedSingleGraphene2023a}. However, the second-order correlation function ($g^{(2)}(\tau)$) at zero delay is flat, showing these spots do not arise from a single \ac{GQD} (see \refsupfig{g2 mapb}). Importantly, since the \ac{GQD}s were initially well dispersed in the \ac{THF} solution, this result demonstrates that the \ac{GQD}s tend to form small clusters on the surface of \mapb{} during the deposition process. This observation is in strong contrast with the situations where \ac{GQD}s are dispersed in polystyrene or deposited on a hexagonal boron nitride substrate. Indeed, in those cases, single \ac{GQD}s are observed (\refsupfig{hBN}). 

\begin{figure}
    \centering
     \begin{subcaptiongroup}
        \subcaptionlistentry{Confocal fluorescence map}
        \labfig{confocal raster scan}
        \subcaptionlistentry{PL spectrum vs. time}
        \labfig{pl spectrum map}
        \subcaptionlistentry{PL spectra downsampled}
        \labfig{PL spectra downsampled}
    \end{subcaptiongroup}
    \includegraphics[width=\linewidth]{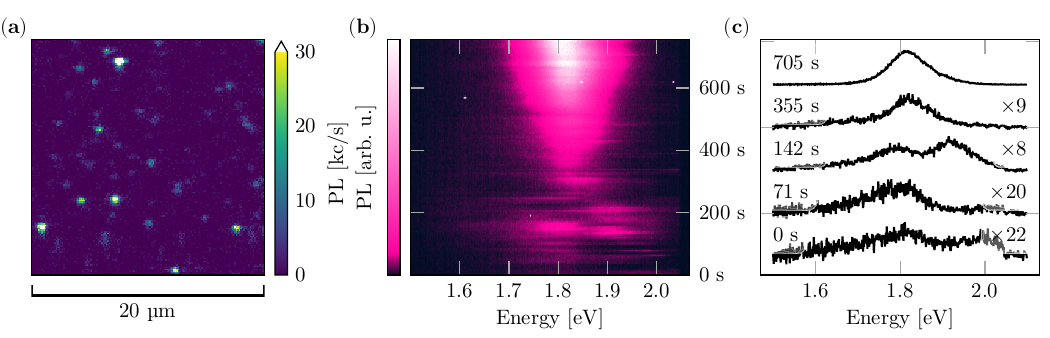}
    \caption{\subreffig{confocal raster scan}~Confocal raster scan of \gqd{}s drop-casted onto an \mapb{} millimetric crystal. \subreffig{pl spectrum map}~\ac{PL} spectra taken repeatidly over \SI{5}{s} exposure time. \subreffig{PL spectra downsampled}~Highlight some of the spectra from \subreffig{pl spectrum map}, showing the 'dancing' dynamics of the emission spectrum. The excitation was set to $2.18~\text{eV}$, with an intensity of $15.47~\text{kW}/\text{cm}^2$.}
    \label{fig:map}
\end{figure}

One example of fluorescence spectroscopy performed on such a cluster is displayed in \reffig{pl spectrum map}. There, the emission spectrum of \gqd{}s on \mapb{} under continuous illumination is repeatedly recorded with an exposure time of \SI{5}{s}. First, we observe the spectra going back and forth ("dancing") between a standard \ac{GQD}-like spectrum with a main line at \SI{1.98}{eV} and one close to the emission energy of the 0-1 vibronic transition of the monomer ($\sim\SI{1.85}{eV}$). Finally, it evolves towards a spectrally-stable redshifted state ($\sim\SI{1.81}{eV}$) with an increased fluorescence intensity. In addition, we observe a dispersion of behaviors from cluster to cluster. Some show the same trend as the one reported in Figure~\ref{fig:map} (see \refsupfig{dancing suppl 1}). Some reach the stabilized high-intensity state faster (see \refsupfig{dancing suppl 2}) and some only show the "dancing" part and stabilize without reaching the high-intensity state (see \refsupfig{dancing suppl 3}).

First, we discuss the "dancing" spectral dynamics. The swing between the two emission lines weighs in favor of a fluctuation between two emissive states rather than a denaturation of the \ac{GQD}s. As the second-order correlation measurements show that the \ac{PL} does not result from a single \ac{GQD}, the "dancing" process could be related to a dynamical coupling-decoupling effect between \ac{GQD}s in a cluster. In particular, the spectrum of the coupled state corresponds well to the one observed on the films (see~\reffig{pl vs concentration}). Therefore, we can interpret the "dancing" spectral dynamics as a consequence of the jump between an uncoupled close to monomer-like state and an excimer state due to the movements of the \ac{GQD}s inside the cluster. Moreover, this type of behavior has already been reported in other molecular systems. For instance, a "dancing" process has been reported in co-facial \ac{PDI} derivatives~\cite{yooExcimerFormationDynamics2010} or in dendrimers functionalized with several chromophores~\cite{Hofkensdancingexcimer}. In their study, Hofkens \textit{et al}~\cite{Hofkensdancingexcimer} designed dendrimers functionalized with several chromophores. Dendrimers are embedded in a polymer film and studied by single-molecule spectroscopy. They observe back-and-forth spectral jumps on a time scale of hundreds of seconds. The spectra arise either from isolated chromophores or from coupled ones. The emission spectra of the coupled chromophores are diverse, globally showing a redshift and modifications of the spectral shape. Likewise, Yoo \textit{et al}~\cite{yooExcimerFormationDynamics2010} synthesized \ac{PDI} molecules linked by xanthene groups. The linkers force the \ac{PDI} molecules to stand face to face, increasing their probability of coupling together. They study the behavior of dimers and trimers. In solution, they observe a redshift of the emission spectrum compared to the monomer case, which they attributed to the coupling between \ac{PDI} molecules in an excimer-like state. Then they perform time-resolved \ac{PL} experiments as a function of time at the single molecule level. A lifetime fluctuation is attributed to a change in the conformation of the dimer and the trimer with time that leads to modifications of the intermolecular distance and relative orientation, and then to a modulation of the coupling. These fluctuations are observed on the time scale of a few tens of seconds. Therefore, we interpret the temporal modifications of the spectrum as a consequence of back-and-forth jumps between coupled and decoupled states due to the movement of \ac{GQD}s in a cluster.

\begin{figure}
    \centering
    \includegraphics{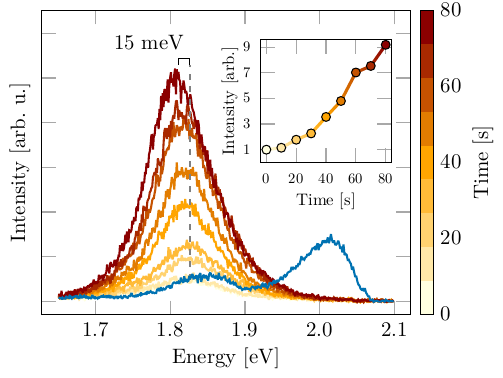}
    \caption{\ac{PL} spectra taken repeatedly over a ten-second exposure. The sample was excited at \SI{2.18}{eV}, with an intensity of \SI{3}{kW\per\cm^2}. The \textcolor{color_gqd_solo}{solid blue line} shows the reference \ac{PL} spectrum of a single \gqd{} in polystyrene. The intensity of the reference spectrum is arbitrary and should not be compared to the other spectra. The inset shows the evolution of the integrated intensity over time.}
    \label{pl increase}
\end{figure}

We discuss now the transition to the redshifted bright state. Figure~\ref{pl increase} shows that the growth of the \ac{PL} signal is accompanied by an additional redshift of the main emission peak by approximately \SI{15}{meV} (relative to the 0-1 transition energy). The inset of Figure~\ref{pl increase} shows an increase by a factor of $\sim$10 of the integrated intensity. This factor typically ranges between 8 and 20, depending on the cluster. We performed time-resolved fluorescence experiments to get insight into the nature of these different states. In particular, using \ac{TTTR} measurements, we recorded the lifetimes associated with two distinct energy ranges of the emission spectrum, as underlined in \reffig{TRPL decay filter}. Overall, the \ac{PL} decay of the integrated spectrum is bi-exponential (black data points). The fit gives two lifetimes, one of \qty{0.65}{ns}, which is close to the instrument response time, and one of \qty{2.65}{ns}. It is in stark contrast with the mono-exponential decay observed in solution (\SI{2.9}{ns}, see \refsupfig{trpl solution}) or at the single-molecule level in polystyrene (\SI{2.41}{ns}, see \refsupfig{trpl ps}). \reffig{TRPL decay filter} displays \ac{PL} decay curves measured for photons emitted respectively in the spectrum's low and high energy ranges. Here, collected photons are detected with a specific bandpass filter before the avalanche photodiode. The energy-resolved \ac{TRPL} curves demonstrate that the long decay component is associated with the high-energy emission range (around the 0-0 emission), while the fast decay component corresponds to lower emission energies (below the energy of the 0-1 transition). Therefore, it is possible to conclude that the single close to \ac{GQD}-like emission spectrum is associated with a decay time of \qty{2.65}{ns} that is close to the one measured in solution and on a single \ac{GQD} in a polystyrene matrix. In comparison, the redshifted spectrum shows a shorter decay time of \qty{0.65}{ns}, although we note that statistically, the shortest lifetime can vary about \SI{0.5}{ns} around this value as shown in \refsupfig{trpl_700nm}. Interestingly, \reffig{TRPL of time} shows how the \ac{PL} intensity of the system integrated over the whole spectrum can evolve over a minute time scale from a state leading to a relatively long mono-exponential decay towards a state resulting in a bi-exponential decay (extra short component). The superimposition of the decay traces taken before and after the dynamical transition is even more striking (see inset of \reffig{TRPL of time}). In particular, we stress that these two \ac{PL} decay curves are not normalized in intensity, which shows that the absolute contribution of the slow decay rate remains mostly unchanged when the system undergoes the transition. In contrast, the short decay component correlates with higher intensities. We observed several cases of this behavior under constant laser illumination (see \refsupfig{additional danse trpl}).

\begin{figure}
    \centering
    \begin{subcaptiongroup}
        \subcaptionlistentry{TRPL decay filter}
        \labfig{TRPL decay filter}
        \subcaptionlistentry{TRPL of time}
        \labfig{TRPL of time}
    \end{subcaptiongroup}
    \includegraphics{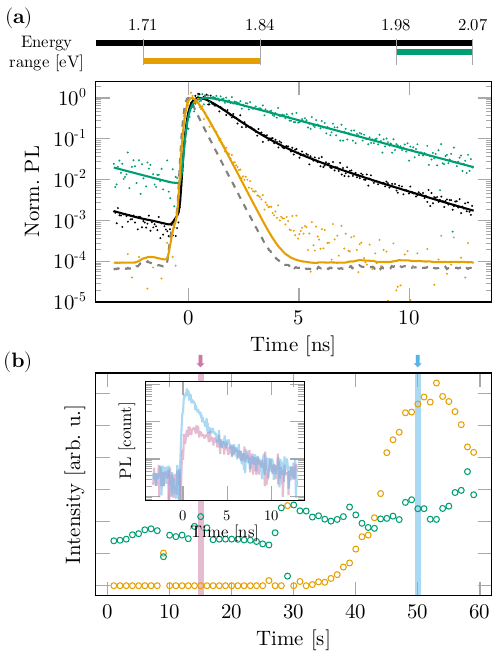}
    \caption{\subreffig{TRPL decay filter}~Comparison of the \ac{TRPL} decay curves for photons integrated over the high-energy range of the spectrum (in \textcolor{color_trpl_high_energy}{green}), the low-energy range (in \textcolor{color_trpl_low_energy}{orange}), and over the whole spectrum (in black). The specific energy ranges are explicit in the upper part of the figure. The baseline of experimental data has been subtracted, see \nameref{sec:methods} and \refsupfig{explanation fitting procedure} for details of the fitting procedure. The \ac{IRF} is plotted in dashed grey. The fitted mono-(for experiments with filters) and bi-(for the whole spectrum) exponential decays are shown with solid lines. \subreffig{TRPL of time}~Evolution of the intensity of the fast  (in \textcolor{color_trpl_low_energy}{orange}) and of the slow component (in \textcolor{color_trpl_high_energy}{green}) of the decay of the integrated \ac{PL} over time. The components are fitted over $1~\text{s}$ integrated decays. The inset highlights the un-normalised decays at $t=15~\text{s}$ (in \textcolor{color_trpl_samples_1}{purple}) and $t=50~\text{s}$ (in \textcolor{color_trpl_samples_2}{light blue}).}
    \label{fig:trpl}
\end{figure}

A consequence of the \ac{GQD}s' mobility would be an alignment of some them, for instance, along the polarization direction of the excitation laser. It could result in a partial organization of the \ac{GQD}s in the cluster. We tried to test this possibility by performing polarization-dependent experiments. Indeed, the emission of single \ac{GQD} is strongly linearly polarized along the \ac{GQD} axis\cite{levy-falkInvestigationRodShapedSingleGraphene2023a}. For a single \ac{GQD} in a polystyrene matrix, the extinction ratio between perpendicular polarizations is of the order of ten (see \refsupfig{pola single gqd}). For \ac{GQD} clusters on \mapb{}, a small degree of linear polarization (ratio of the order of two) has been measured (see \refsupfig{pola gqd on mapb}), showing a partial organization of \ac{GQD}s in a given cluster. Nevertheless, we fail to relate it to a dynamical process under illumination. 

Significantly, results from time-resolved \ac{PL} demonstrate that the coupled state has a shorter lifetime combined with a higher \ac{PL} intensity. Together with the additional redshift of the emission, these results are compatible with the emergence of a partial collective emission from coupled excited states within a partially organized \ac{GQD} cluster~\cite{collectiveMePTCDI}. Since this kind of collective state depends on the distance and relative orientation between emitters, the time evolution of the optical properties suggests a modification of the organization of \ac{GQD}s and intermolecular distances in the clusters under illumination. However, the driving mechanism and special effect of the perovskite substrate remain unclear.

\section{Conclusion}

We investigated the photophysics of \gqd{} graphene quantum dots deposited on a perovskite substrate. When the density is high, \ac{GQD}s form an inhomogeneous film where the fluorescence shows an excimer characteristic with a spectrum centered close to the 0-1 vibronic line. When the concentration of \ac{GQD}s is decreased, and despite observing diffraction-limited \ac{PL} spots, the samples show features readily attributed to the interaction between \ac{GQD}s in aggregates. Under laser excitation, we observe the \ac{GQD}s jumping from electronically coupled to uncoupled states. Furthermore, when the sample is kept under constant illumination, the system can reach a stable state with a slightly more redshifted emission correlated with an increase in the \ac{PL} intensity and a shorter lifetime. We attribute this later effect to a reorganization of the \ac{GQD}s at the surface of the perovskite, leading to an emitting collective excited state. 

The driving force of these effects remains unclear. Several parameters will be varied to get more insights into the mechanisms at play. For instance, lowering the temperature to a few Kelvin will give valuable information about the dynamical coupling-decoupling process since the mobility of the \ac{GQD}s on the surface will be modified. Likewise, lowering the number of \ac{GQD}s in a cluster might help better understand and control the process. A detailed understanding of the organization of the \ac{GQD}s on-surface would open the possibility of building arrays of coherently coupled emitters. 

\begin{backmatter}

\bmsection{Funding}
The authors acknowledge ﬁnancial support from the Ministry of Armies (Agence de l'innovation de défense). This work was ﬁnancially supported by the FLAG-ERA Grant OPERA by DFG 437130745 and ANR-19-GRF1-0002-01, by the ANR-DFG NLE Grant GRANAO ANR-19-CE09-0031-01, ANR grant DELICACY ANR-22-CE47-0001-03, ANR grant GANESH ANR-21-CE09-0025, and by a public grant overseen by the French National Research Agency (ANR) as part of the "Investissements d'Avenir" program (Labex NanoSaclay, reference: ANR-10-LABX-0035).

\bmsection{Disclosures}
The authors declare no conflicts of interest.

\bmsection{Data availability} Data underlying the results presented in this paper are not publicly available at this time but may be obtained from the authors upon reasonable request.

\bmsection{Supplemental document}See Supporting Information for Experimental Setup, \mapb{} substrate characterization, Additional spectroscopic characterizations of \gqd{}.

\end{backmatter}
\bibliography{bibliography}

\end{document}

\renewcommand{\thesection}{S.\arabic{section}}
\renewcommand{\thefigure}{S\arabic{figure}}

\title{Supplementary materials: Collective Fluorescence of Graphene Quantum Dots on a surface}

\author{Hugo Levy-Falk,\authormark{1,2,$\dag$} Suman Sarkar,\authormark{1,$\dag$} Thanh Trung Huynh,\authormark{1} Daniel Medina-Lopez,\authormark{3} Lauren Hurley,\authormark{1} Océane Capelle,\authormark{1} Muriel Bouttemy,\authormark{4} Gaëlle Trippé-Allard,\authormark{1} Stéphane Campidelli,\authormark{3} Loïc Rondin,\authormark{1} Elsa Cassette,\authormark{1} Emmanuelle Deleporte,\authormark{1} Jean-Sébastien Lauret\authormark{1,*}}

\address{\authormark{1}Laboratoire LuMIn, Université Paris-Saclay, ENS Paris-Saclay, CentraleSupélec, CNRS, LuMIn, 91190 Gif-sur-Yvette, France\\
\authormark{2}National Institute of Optics (CNR-INO), c/o LENS via Nello Carrara 1, Sesto F.no 50019, Italy\\
\authormark{3}Université Paris Saclay, CEA, NIMBE, LICSEN, \textsc{Gif-sur-Yvette, France}\\
\authormark{4}Institut Lavoisier de Versailles, UVSQ, Université Paris-Saclay, CNRS, UMR 8180, 45 avenue des Etats-Unis, 78035 Versailles Cedex, France\\
\authormark{$\dag$}The authors contributed equally to this work.}

\email{\authormark{*}lauret@ens-paris-saclay.fr}

\section{Characterization of the \mapb{} substrate}
\label{sec:characterization of mapb}

\begin{figure}[H]
    \centering
    \includegraphics{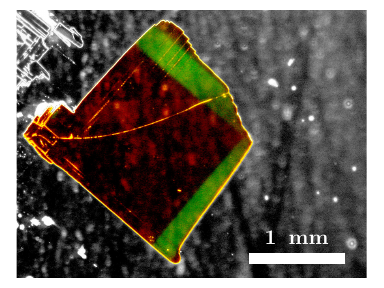}
    \caption{Optical microscopy picture of a halide perovskite crystals used as a substrate for this study. The background has been grayed out. The two visible green bands come from the photoluminescence of the crystal excited by the LEDs of the microscope. The shape of these fluorescence bands are due to the specific orientation of the lamps.}
    \labfig{optical microscope}
\end{figure}
\begin{figure}[H]
  \centering
  \includegraphics{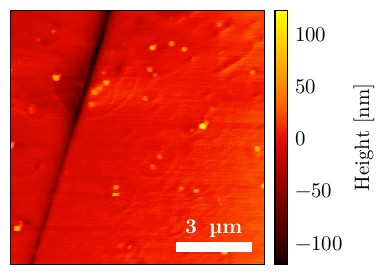}
    \caption{Atomic Force Microscopy map of the surface of \mapb{} crystals. The surface roughness is \SI{18}{nm}, and \SI{10}{nm} when removing the crevice from the measured area.}
  \labfig{afm}
\end{figure}

\subsection{XPS study}

In the case of perovskites, specific care has to be taken to ensure the reliability of measurements in UHV and under X-rays irradiation. Even if we expect that the bromine-based perovskite substrates developed are more stable than iodine-basedperovskites, a similar analytical methodology, thoroughly described in Ref.~\cite{cacovichLightInducedPassivationTriple2020} has been used.
The samples, consisting in millimeter-sized crystals are enclosed between glass coverslips for transportation. After separating the glass coverslips, one of them was mounted on the sample holder, leading to a brief exposure to ambient air before the introduction inside the spectrometer, NEXSA XPS system from Thermo Fischer Scientific, using a monochromatic Al K$\alpha$ X-ray source (1486.6 eV). The source power selected for the measurements was 72 W (12 kV, 60 mA). Charge compensation was necessary, leading to a chamber pressure was approximately $\SI{3e{-7}}{mbar}$. The beam size was set at $\SI{300}{\micro\meter} \times \SI{600}{\micro\meter}$ to best match the size of the crystals. Avantage software was used for data treatment. Quantification was done by employing a Shirley background and RSF from the library of the constructor. Chemical environment differentiation was realized using contributions with L/G of \SI{30}{\percent}. 

\begin{figure}
    \centering
    \includegraphics[width=0.5\linewidth]{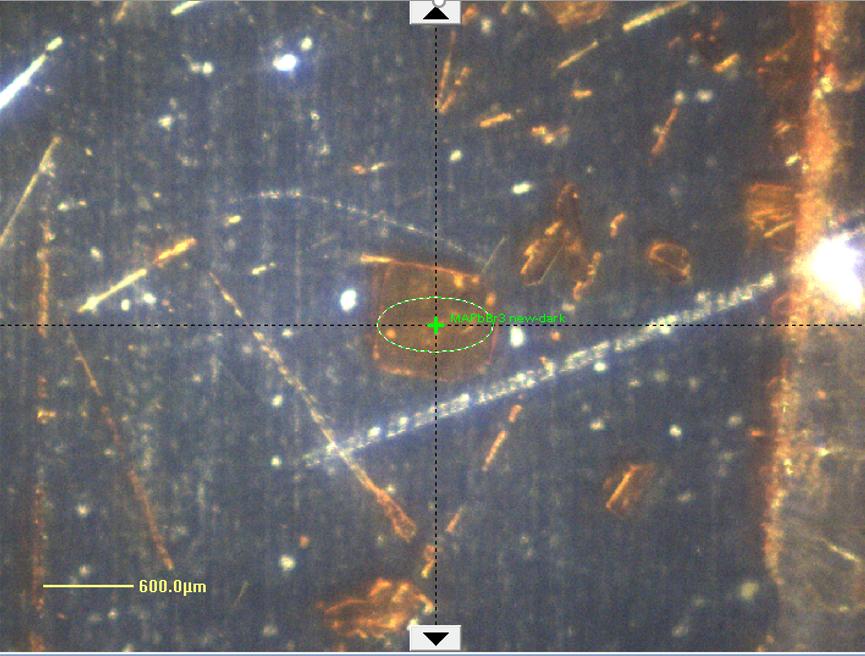}
    \caption{IR camera view of the sample inside the analysis chamber and position of the $\SI{300}{\micro\meter} \times \SI{600}{\micro\meter}$ analysis X-ray spot size on a \mapb{} crystal. }
    \labfig{photo xps chamber}
\end{figure}

A typical survey spectrum is shown in \refsupfig{xps survey}, featuring the expected elments C, N, O, Br, and Pb. High-energy resolution C 1s, N 1s, O 1s, Br 3d and Pb 4f photopeaks and two surveys were acquired, at the beginning and the end of the experiment, showing a slight atomic reordering under X-ray action (total acquisition time of 46 min). The reproducibility of the measurement and homogeneity of the composition was verified on three different crystals.

\begin{figure}[H]
  \centering
  \begin{subcaptiongroup}
      \subcaptionlistentry{xps survey}
      \labfig{xps survey}
      \subcaptionlistentry{xps survey br}
      \labfig{xps survey br}
      \subcaptionlistentry{xps survey c}
      \labfig{xps survey c}
      \subcaptionlistentry{xps survey N}
      \labfig{xps survey N}
      \subcaptionlistentry{xps survey O}
      \labfig{xps survey O}
      \subcaptionlistentry{xps survey Pb}
      \labfig{xps survey Pb}
  \end{subcaptiongroup}
  \includegraphics[width=\linewidth]{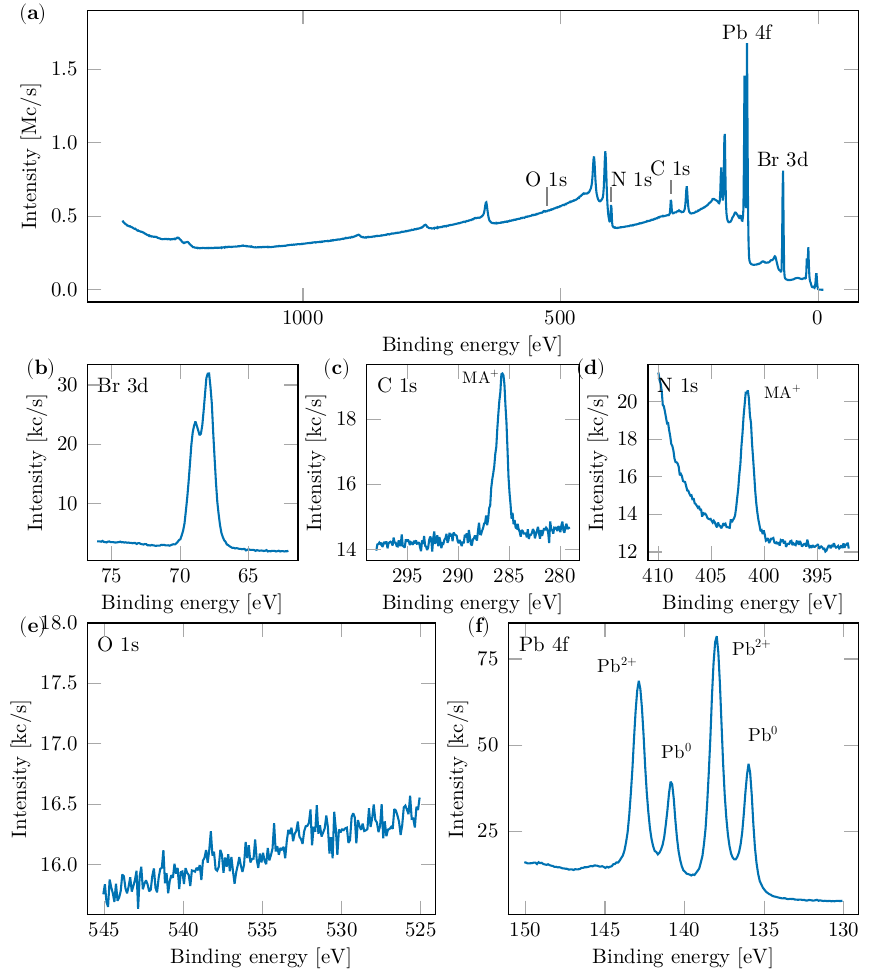}
    \caption{\subreffig{xps survey}~XPS survey spectrum measured on the perovskite’s surface with identification of the main transitions of the constitutive elements. The specific core levels tracked are shown in high energy resolution in \subreffig{xps survey br}~for Br 3d, \subreffig{xps survey c}~for C 1s, \subreffig{xps survey N}~for
N 1s, \subreffig{xps survey O}~for O 1s, and \subreffig{xps survey Pb}~for Pb 4f.}
  \labfig{xps big}
\end{figure}

The atomic percentages obtained are shown in \refsuptab{xps proportion}. The peak positions of C at \SI{285.8(0.2)}{eV}, N at \SI{401.5(0.2)}{eV}, Pb 4f$_{7/2}$ at \SI{137.9(0.2)}{eV}, and Br at \SI{67.8(0.2)}{eV} are consistent with the ones expected~\cite{bechuPhotoemissionSpectroscopyCharacterization2020} if we consider a shift of 1 eV towards lower energies, which we attribute to surface potential effects. Strikingly, the amount of oxygen is below the detection limit, as shown in \refsupfig{xps survey O}. The amount of carbon is also low, and can be mostly attributed to \ch{MA+}, as shown in the fits in \refsupfig{xps}. In this figure, the main peak is associated to C 1s MA+ two other contributions are required to model the C 1s signal and corresponds to the superficial carbon contamination which is accounted for by carbons presenting C-C or C-O bonds.

\begin{table}[H]
  \centering
  \begin{tabular}{ccccc}
    \hline
      \textbf{C} & \textbf{Br} & \textbf{N} & \textbf{O} & \textbf{Pb} \\
    \hline
      \SI{14.4}{\%} & \SI{62.3}{\%} & \SI{13.1}{\%} & \SI{0.0}{\%} & \SI{10.2}{\%} \\
    \hline
  \end{tabular}
  \caption{XPS atomic percentages measured on the bare substrate. Details on the procedure to find atomic percentages can be found in Ref.~\cite{vidonImpactXRayRadiation2023}}
  \labtab{xps proportion}
\end{table}

\begin{figure}[H]
  \centering
    \includegraphics{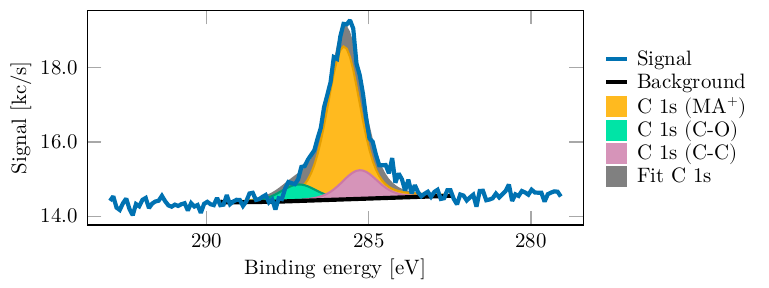}
    \caption{C 1s photopeak at the surface of the \mapb{} sample. 3 different chemical environments associated to MA$^+$ and carbonaceous contamination (C-O and C-C) are represented by the coloured envelopes}
  \labfig{xps}
\end{figure}

The precise decomposition of the C 1s spectrum enables to separate the contribution of the carbonaceous contamination (C-O and C-C) from the contribution of the C-N bonding of MA+. 

Finally, a picture of the composition of the surface can be constructed from the comparison of the measured atomic ratio with their expect ratio, as shown in \refsuptab{xps comparison} (note that \ch{MABr} is present with \SI{10}{\percent} excess in the synthesis). Remarkably, the amount of \ch{MA} is close to the expected value, which validate the fitting procedure in \refsupfig{xps}. On the other hand, the ratio of any species against Br is systematically low, which would indicate a higher amount of Br at the surface of the perovskite.

\begin{table}[H]
  \centering
  \begin{tabular}{ccc}
    \hline
    \textbf{Ratio} & \textbf{Measured}  & \textbf{Expect.} \\
    \hline
    C*/Pb & 1.19 &  1.10 \\
    N/Pb & 1.29 &  1.10 \\
    O/Pb & 0.00 &  0.00 \\
    C*/N & 0.93 &  1.00 \\
    C*/Br & 0.19 & 0.35 \\
    N/Br & 0.21 & 0.35 \\
    Pb/Br & 0.16 & 0.32 \\
    \hline
    C/Pb & 1.41 & 1.10 \\
    C/N & 1.10 & 1.00 \\
    C/Br & 0.23 & 0.35 \\
    \hline
    \ch{Pb^{2+}}/\ch{Pb^0} & 2.04 & \\
    \hline
  \end{tabular}
  \caption{Atomic ratios deduced from atomic proportions given in \reftab{xps proportion}. C* denotes the atomic percentage originating only from \ch{MA+} cations (\textit{i.e.} after subtraction of the C contamination contribution). The second part of the table accounts for all carbon content (including contamination). The third part compares the peaks of ionic and metallic lead. The expected ratios account for a \SI{10}{\%} excess of \ch{MABr} in the synthesis.}
  \labtab{xps comparison}
\end{table}

\section{Synthesis of \gqd{}}

\begin{figure}[H]
    \centering\includegraphics[alt={Synthesis of the GQD}, width=\textwidth]{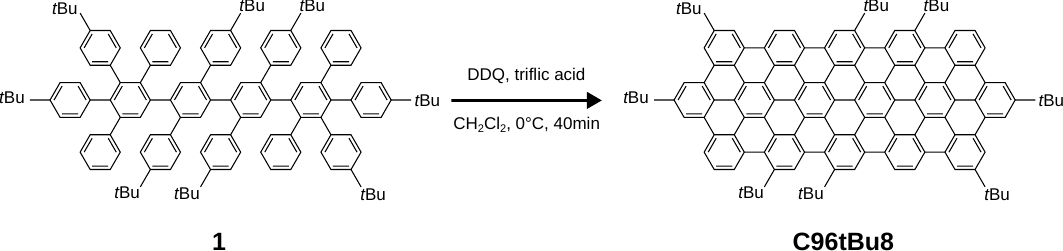}
    \caption{Oxidation of the polyphenylene dendrimer 1 with 2,3-dichloro-5,6-dicyano-p-benzoquinone (DDQ) in the presence of triflic acid in dichloromethane to give the \gqd{} GQD.}
    \labfig{synthesis oxydation}
\end{figure}

The synthesis of the polyphenylene dendrimer \textbf{1} was reported previously~\cite{medina-lopezInterplayStructurePhotophysics2023}. Conversely to our previous paper in which the \textbf{\gqd{}} GQD was prepared by oxidation of dendrimer \textbf{1} using \ch{FeCl3}; here, the dendrimer \textbf{1} was oxidized using DDQ in the presence of triflic acid in dichloromethane.

\textbf{\gqd{}}. Dendrimer \textbf{1} (\SI{30}{mg}, \SI{0.018}{mmol}) and 2,3-dichloro-5,6-dicyano-1,4-benzoquinone (\SI{132}{mg}, \SI{0.58}{mmol}) were dissolved in freshly distilled unstabilized dichloromethane (\SI{15}{mL}). The solution was allowed to cool down to \SI{0}{\degreeCelsius} in an ice bath. Trifluoromethanesulfonic acid (\SI{0.9}{mL}) was slowly added to the mixture. The solution is stirred for \SI{40}{min} under argon at \SI{0}{\degreeCelsius}. Distilled trimethylamine (\SI{5}{mL}) was added to quench the reaction and the mixture was then added to an excess of methanol (\SI{100}{mL}). The precipitate was filtered through a PTFE filter. The obtained powder was dispersed in a small amount of tetrahydrofuran, sonicated for \SI{1}{min} followed by ultracentrifugation (\SI{130 000}{g}, \SI{30}{min}). The precipitate yielded the product as a dark green powder (\SI{11}{mg}, \SI{26}{\percent}). The \textbf{\gqd{}} GQD was characterized by MALDI-TOF mass spectrometry.

\section{Experimental setup}
\begin{figure}[H]
    \centering\includegraphics[width=\textwidth, alt={Schematic of our confocal fluorescence microscope.}]{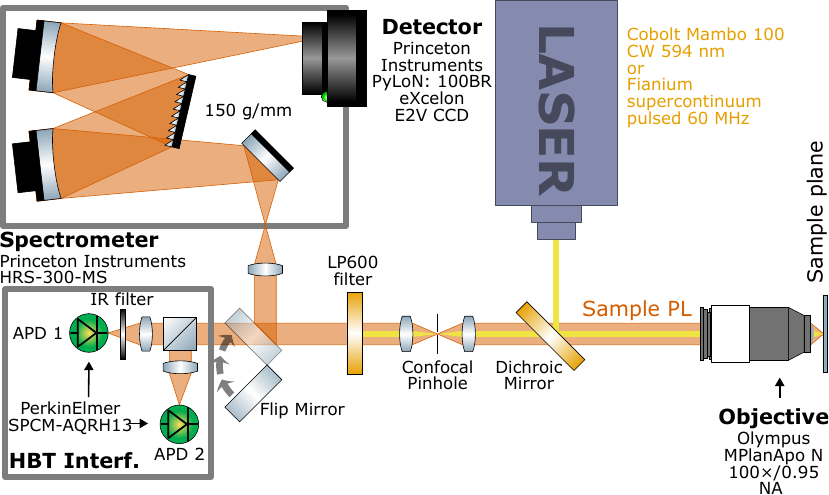}
    \caption{Schematics of the confocal fluorescence microscope. }
    \labfig{confocal microscope}
\end{figure}

\section{Optical properties of \gqd{}}

\begin{figure}[H]
  \centering
  \includegraphics{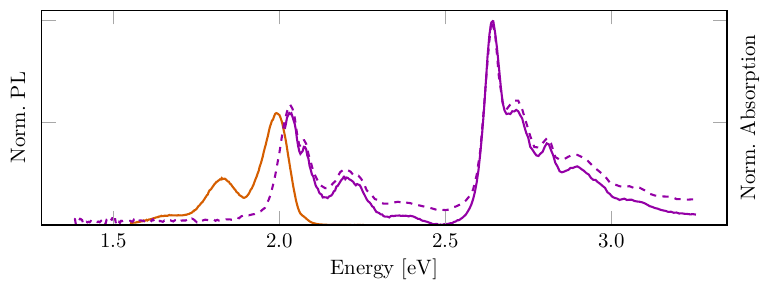}
    \caption{Spectroscopy in solution of \gqd{}. Emission spectrum is in \textcolor{color_pl_gqd_solution}{vermillon}, absorption spectrum is in \textcolor{color_ple_gqd_solution}{dashed purple}, and excitation spectrum in \textcolor{color_abs_gqd_solution}{solid purple}.}
  \labfig{spectro solution}
\end{figure}

\begin{figure}[H]
  \centering
  \includegraphics{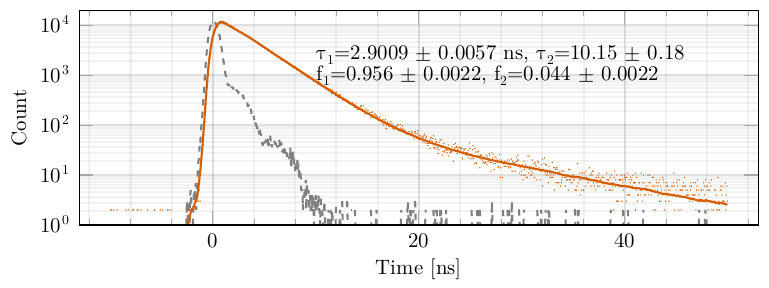}
    \caption{Time-Resolved Photoluminescence of \gqd{} in solution. The instrument response function is the dashed \textcolor{grey}{grey} line, the experimental data are the \textcolor{color_trpl_gqd_solution}{vermillon dots}, and the mono-exponential decay fit is the \textcolor{color_trpl_gqd_solution}{vermillon solid line}. The error bars are given by the fit procedure. Given the very low fractional number of the long decay rate ($<\SI{5}{\percent}$), it is safe to treat it as a monoexponential decay.}
  \labfig{trpl solution}
\end{figure}

\begin{figure}[H]
  \centering
  \includegraphics{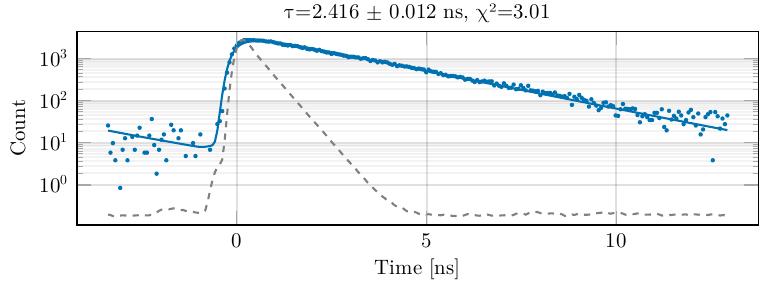}
    \caption{Time-Resolved Photoluminescence decay of a single \gqd{} in the polystyrene matrix. The instrument response function is the \textcolor{grey}{grey} dashed line, the experimental data are the  \textcolor{color_trpl_gqd_poly}{blue dots}, and the mono-exponential decay fit is the \textcolor{color_trpl_gqd_poly}{blue solid line}.}
  \labfig{trpl ps}
\end{figure}

\begin{figure}[H]
  \centering
  \includegraphics{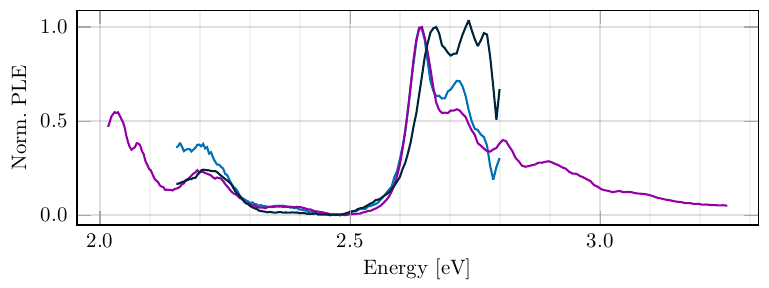}
    \caption{Photoluminescence excitation spectra of \gqd{} in the polystyrene matrix (\textcolor{color_ple_gqd_poly}{blue}), in the solution (\textcolor{color_ple_gqd_solution}{purple}) and of \gqd{} films on \mapb{} substrate (\textcolor{color_ple_gqd_mapb}{dark blue}). The maximum is at \SI{2.64}{eV} for the samples in solution and in the polystyrene matrix and at \SI{2.77}{eV} for the film on \mapb{}.}
  \labfig{ple}
\end{figure}

\begin{figure}[H]
  \centering
  \begin{subcaptiongroup}
      \subcaptionlistentry{g2 ps}
      \labfig{g2 ps}
      \subcaptionlistentry{g2 mapb}
      \labfig{g2 mapb}
  \end{subcaptiongroup}
  \includegraphics{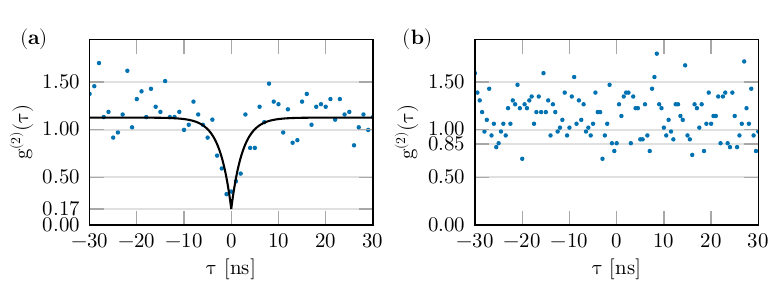}
    \caption{Typical second-order correlation measurement of light emitted by a diffraction-limited spot of \gqd{} \subreffig{g2 ps}~in the polystyrene matrix, and \subreffig{g2 mapb}~on the \mapb{} substrate, excited CW at \SI{594}{nm}. The correlations are fitted with a mono-exponential law, plotted in black.}
  \labfig{g2 in ps}
\end{figure}

\begin{figure}[H]
  \centering
  \begin{subcaptiongroup}
      \subcaptionlistentry{stability trace}
      \labfig{stability trace}
      \subcaptionlistentry{stability stats}
      \labfig{stability stats}
  \end{subcaptiongroup}
  \includegraphics[width=\textwidth]{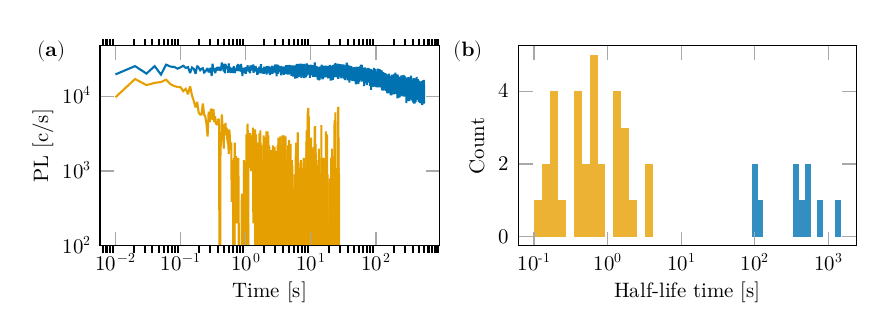}
    \caption{Comparison of the stability of emission of \gqd{} films on glass substrate  (\textcolor{wong_orange}{orange}) and \mapb{} surface (\textcolor{wong_blue}{blue}). \subreffig{stability trace}~Typical time trace of the integrated fluorescence under continuous excitation at \SI{2.09}{eV} \SI{0.396}{kW/cm^2}. \subreffig{stability stats}~Distribution of time required to halve the initial photoluminescence intensity when recording such time traces.}
  \labfig{stability}
\end{figure}

\begin{figure}[H]
  \centering
  \begin{subcaptiongroup}
      \subcaptionlistentry{film on mapb}
      \labfig{film on mapb}
      \subcaptionlistentry{film on glass}
      \labfig{film on glass}
      \subcaptionlistentry{map mapb}
      \labfig{map mapb}
      \subcaptionlistentry{control mapb}
      \labfig{control mapb}
  \end{subcaptiongroup}
  \includegraphics[width=\textwidth]{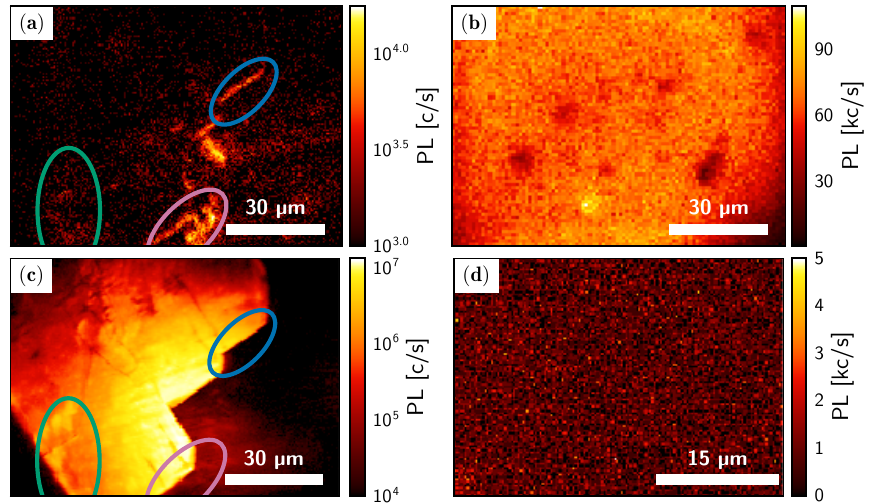}
    \caption{\subreffig{film on mapb}~PL raster scan of \gqd{} film drop-casted on \mapb{} substrate excited at \SI{2.17}{eV}. \subreffig{film on glass}~PL raster scan of a film of \gqd{} in TCB drop-casted on glass substrate. The excitation energy was \SI{2.17}{eV}. \subreffig{map mapb}~PL raster scan of the same area as in \subreffig{film on mapb}, excited at \SI{3.06}{eV}, integrated for energies lower than \SI{2.07}{eV}. \subreffig{control mapb}~Photoluminescence raster scan of a \mapb{} crystal soaked in TCB, then dried for 2h30 at \SI{60}{\degreeCelsius}. Excitation energy was \SI{2.06}{eV}. Scans \subreffig{film on mapb} and \subreffig{map mapb} were excited using the supercontinuum laser (\SI{60}{MHz} repetition rate) whereas scan \subreffig{film on glass} was excited using our continuous wave Cobolt laser.}
  \labfig{substrate and film characterization}
\end{figure}

\begin{figure}[H]
  \centering
  \begin{subcaptiongroup}
      \subcaptionlistentry{raster film on mapb}
      \labfig{raster film on mapb}
      \subcaptionlistentry{raster film on mapb diff limited}
      \labfig{raster film on mapb diff limited}
      \subcaptionlistentry{raster mapb diff limited}
      \labfig{raster mapb diff limited}
  \end{subcaptiongroup}
  \includegraphics[width=\textwidth]{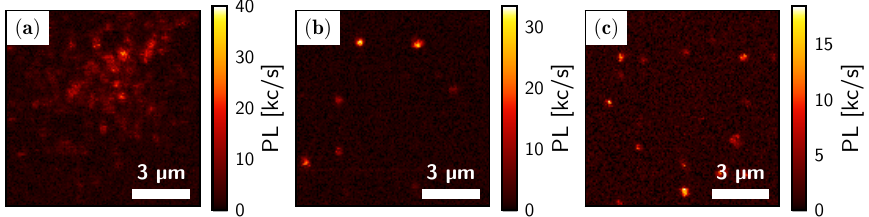}
    \caption{\subreffig{raster film on mapb}~Typical photoluminescence raster scan of \gqd{} on \mapb{}, spin-coated from TCB solution. Excitation intensity was \SI{0.54}{kW/cm^2}. \subreffig{raster film on mapb diff limited}~Photo-luminescence raster scan of \gqd{} on \mapb{}, spin-coated from TCB solution, featuring diffraction-limited emission spots. Excitation intensity was \SI{0.26}{kW/cm^2}. \subreffig{raster mapb diff limited}~Typical photoluminescence raster scan of \gqd{} on \mapb{}, spin-coated from THF solution. Excitation intensity was \SI{0.26}{kW/cm^2}.}
  %\labfig{substrate and film characterization}
\end{figure}

\section{h-BN substrate}
\begin{figure}[H]
    \centering
    \begin{subcaptiongroup}
      \subcaptionlistentry{hBN microscopy}
      \labfig{hBN microscopy}
      \subcaptionlistentry{hBN raster scan}
      \labfig{hBN raster scan}
      \subcaptionlistentry{hBN g2}
      \labfig{hBN g2}
  \end{subcaptiongroup}\includegraphics[width=\textwidth]{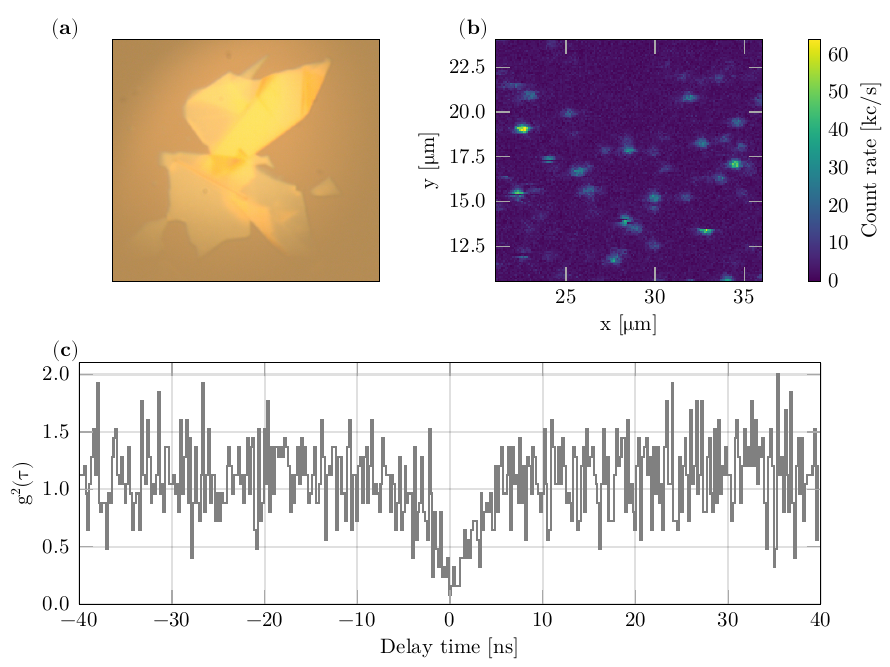}
    \caption{\subreffig{hBN microscopy}~Optical image of exfoliated h-BN flakes. \subreffig{hBN raster scan}~Photoluminescence raster scan of GQD (C$_{114}$\textit{t}Bu$_8$) on h-BN flake. \subreffig{hBN g2}~Second-order correlation function showing g$^2$(0)$<0.5$.}
    \labfig{hBN}
\end{figure}
\section{Dancing spectra}
\begin{figure}[H]
    \centering
    \includegraphics[width=\textwidth]{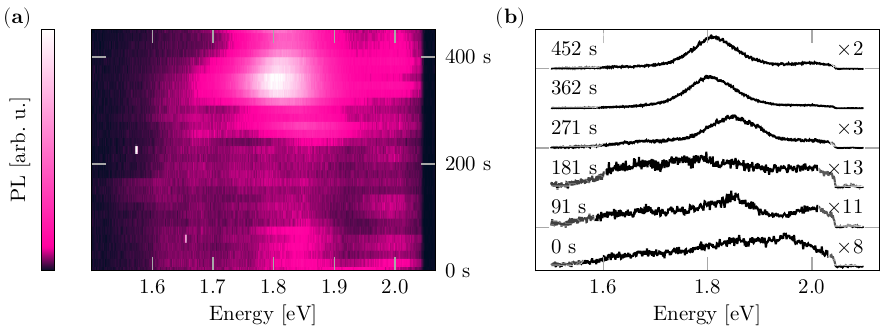}
    \caption{"Dancing spectrum" of \gqd{} on \mapb{}, showing a violent increase of the fluorescence intensity in the red after 300 seconds. Illumination power was \SI{15.47}{\kilo\W\per\cm^2} at \SI{2.17}{eV}. }
    \labfig{dancing suppl 1}
\end{figure}
\begin{figure}[H]
    \centering
    \includegraphics[width=\textwidth]{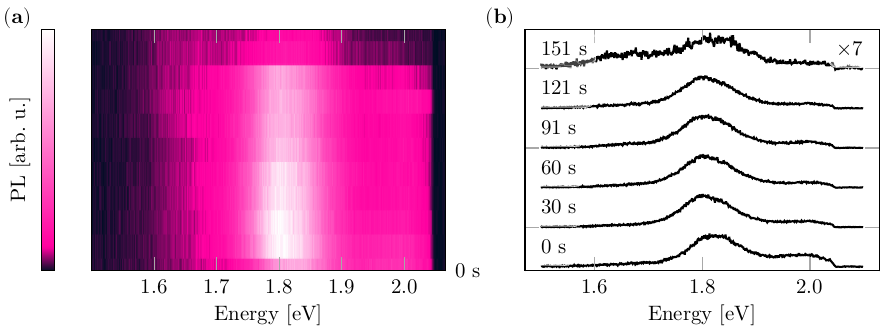}
    \caption{"Dancing spectrum" of \gqd{} on \mapb{}, showing a violent increase of the fluorescence intensity in the red after 10 seconds. Illumination power was \SI{15.47}{\kilo\W\per\cm^2} at \SI{2.17}{eV}.}
    \labfig{dancing suppl 2}
\end{figure}
\begin{figure}[H]
    \centering
    \includegraphics[width=\textwidth]{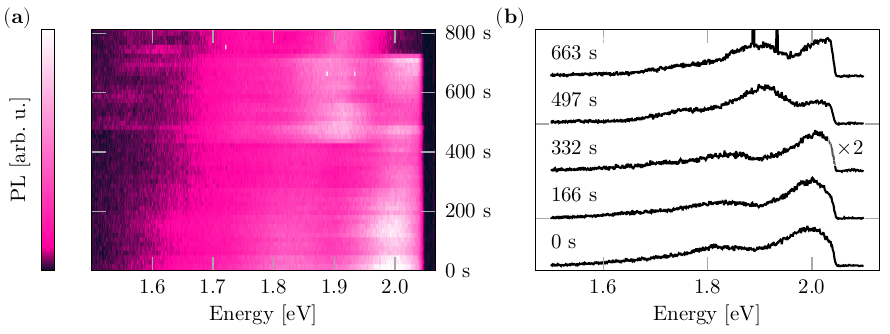}
    \caption{"Dancing spectrum" of \gqd{} on \mapb{}, showing back-and-forth movement between "normal" and "red-shifted" fluorescence spectra. Illumination power was \SI{15.47}{\kilo\W\per\cm^2} at \SI{2.17}{eV}.}
    \labfig{dancing suppl 3}
\end{figure}

\section{Statistics of TRPL}

\begin{figure}[H]
    \centering
    \includegraphics[width=\textwidth]{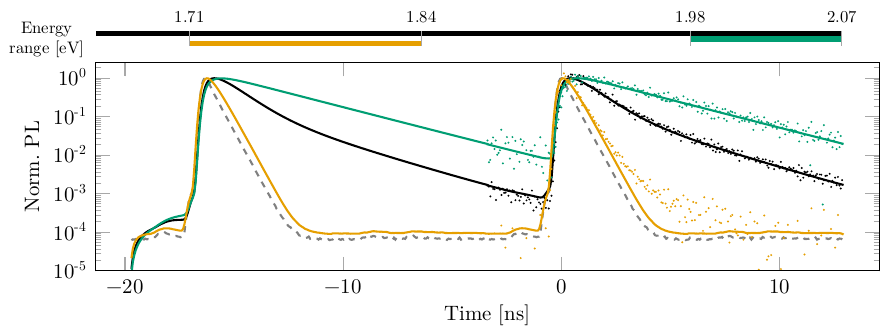}
    \caption{Illustration of the fitting procedure for TRPL decays. The IRF is shown in dashed gray and is reconstructed to include a prepulse. The filtered high energy decay is in \textcolor{color_trpl_high_energy}{green}, the filtered low-energy decay is in \textcolor{color_trpl_low_energy}{orange}, and the decay integrated over the whole spectrum is in black. The specific energy ranges are explicited in the upper part of the figure. The (bi-)exponential decays are fitted to include the pre-pulse, as shown here. The baseline of the experimental data has been subtracted.}
    \labfig{explanation fitting procedure}
\end{figure}

\begin{figure}
    \centering
    \includegraphics[width=\textwidth]{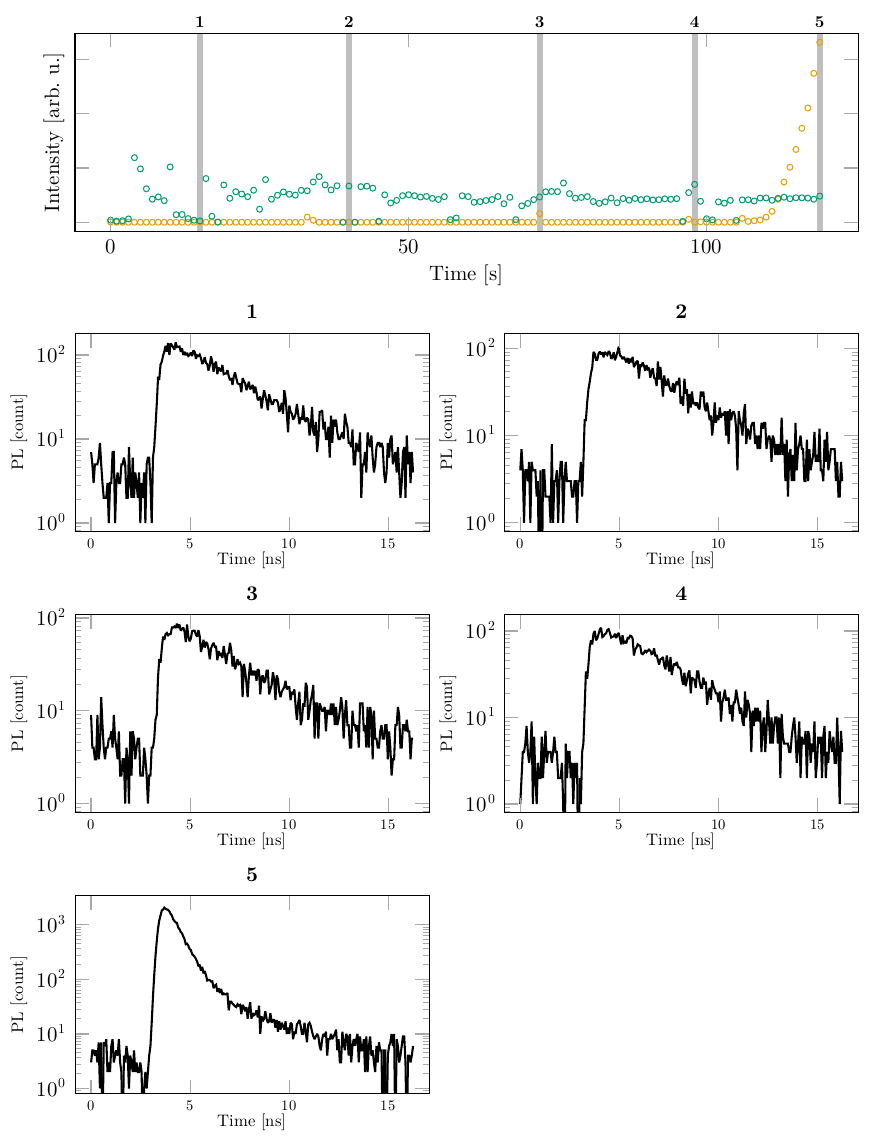}
    \caption{Evolution of the fractional part of the fast component (in \textcolor{color_trpl_high_energy}{orange}) and of the slow component (in \textcolor{color_trpl_low_energy}{green}) of the decay of the integrated photoluminescence over time. The components are fitted over $1~\text{s}$ integrated decays. Individual TRPL traces have been singled out and plotted below.}
    \labfig{additional danse trpl}
\end{figure}

\begin{figure}[H]
  \centering
  \begin{subcaptiongroup}
      \subcaptionlistentry{tau per sample 600}
      \labfig{tau per sample 600}
      \subcaptionlistentry{tau 600}
      \labfig{tau 600}
  \end{subcaptiongroup}
  \includegraphics[width=\textwidth]{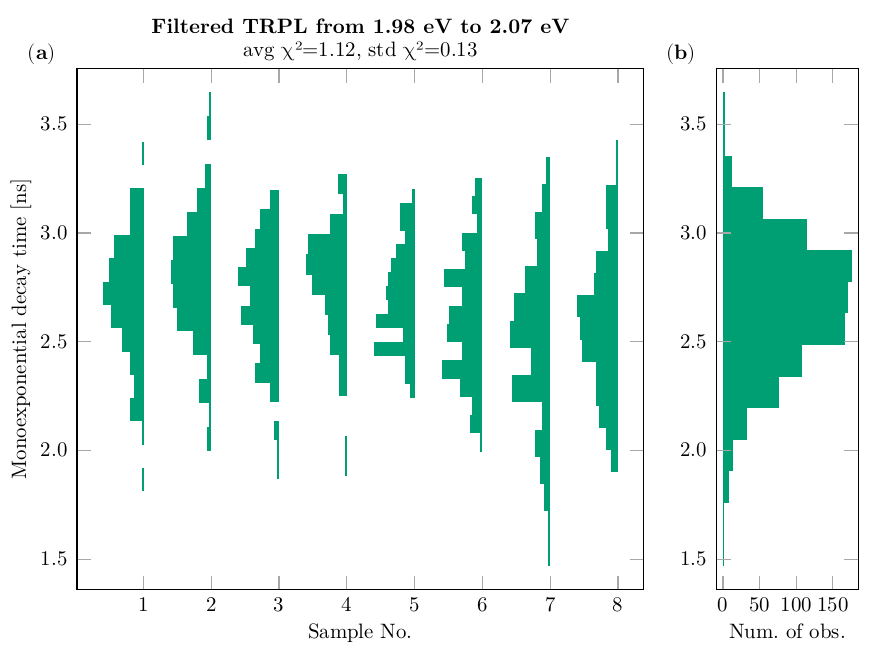}
    \caption{Statistics for TRPL traces with a band-pass filter for the \SI{1.98}{eV}-\SI{2.07}{eV} region. \subreffig{tau per sample 600}~Distribution of lifetime per sample. \subreffig{tau 600}~Overall lifetime distribution.}
  \labfig{trpl_600nm}
\end{figure}

\begin{figure}[H]
  \centering
  \begin{subcaptiongroup}
      \subcaptionlistentry{tau per sample red}
      \labfig{tau per sample red}
      \subcaptionlistentry{tau red}
      \labfig{tau red}
  \end{subcaptiongroup}
  \includegraphics[width=\textwidth]{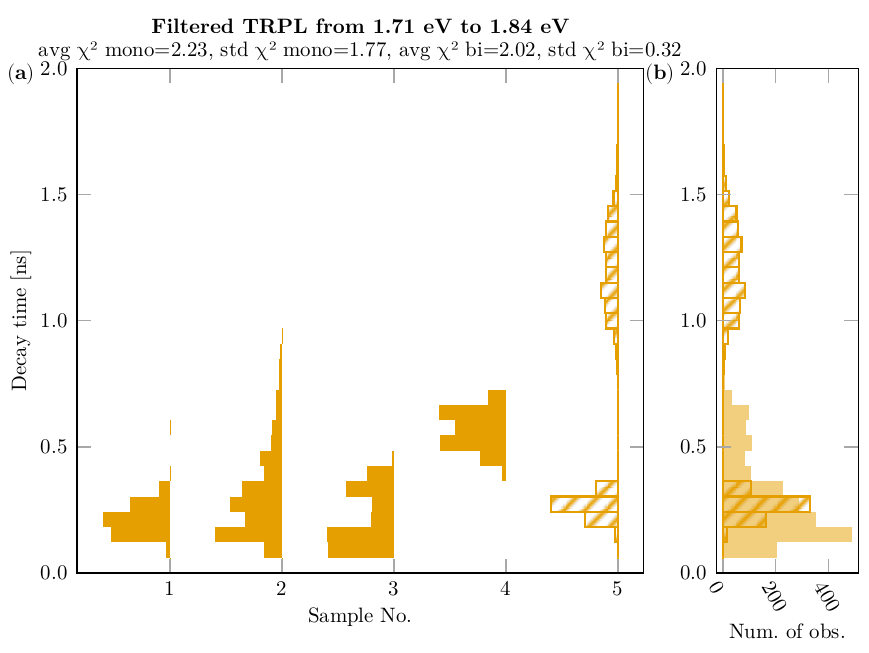}
    \caption{Statistics for TRPL traces with a band-pass filter for the \SI{1.71}{eV}-\SI{1.84}{eV} region. \subreffig{tau per sample red}~Distribution of lifetime per sample. The patterned histograms correspond to bi-exponential decays. \subreffig{tau red}~Overall lifetime distribution, using the same color code.}
  \labfig{trpl_700nm}
\end{figure}
\begin{figure}[H]
  \centering
  \begin{subcaptiongroup}
      \subcaptionlistentry{tau per sample other}
      \labfig{tau per sample other}
      \subcaptionlistentry{tau other}
      \labfig{tau other}
  \end{subcaptiongroup}
  \includegraphics[width=\textwidth]{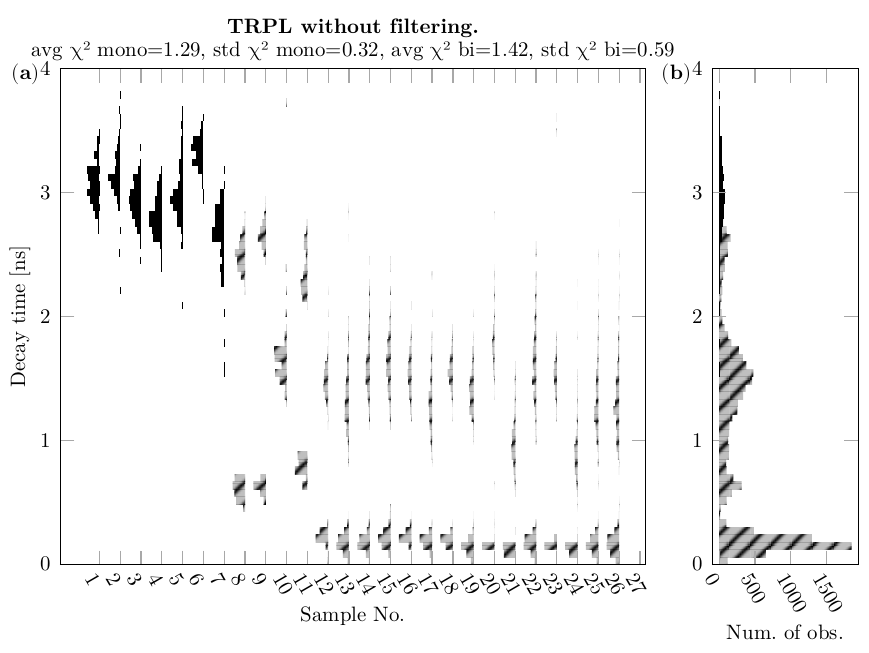}
    \caption{Statistics for TRPL traces without filter. \subreffig{tau per sample other}~Distribution of lifetime per sample.  The patterned histograms correspond to bi-exponential decays. \subreffig{tau other}~Overall lifetime distribution, using the same color code.}
  \labfig{trpl_other}
\end{figure}

\section{Degree of polarization}

\begin{figure}[H]
  \centering
  \begin{subcaptiongroup}
      \subcaptionlistentry{pola single gqd}
      \labfig{pola single gqd}
      \subcaptionlistentry{pola gqd on mapb}
      \labfig{pola gqd on mapb}
  \end{subcaptiongroup}
  \includegraphics[width=\textwidth]{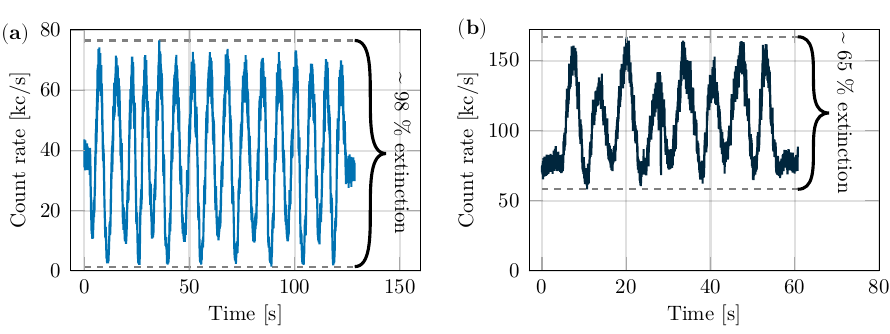}
    \caption{Typical variation of the PL intensity as a function of time while varying by hand the angle of a $\lambda$/2 waveplate positioned in the excitation path. \subreffig{pola single gqd}~Single GQD in a polystyrene matrix. \subreffig{pola single gqd}~Cluster of GQDs on MAPB.}
  \labfig{pola}
\end{figure}

\pagebreak

\bibliography{bibliography}